\documentclass[authoryear,review,11pt]{elsarticle}
\usepackage{graphicx}
\usepackage{appendix}
\usepackage{amsmath,amsfonts}
\usepackage[nonumberlist]{glossaries}
\makeglossaries

\journal{Journal}

\begin{document}
\newglossaryentry{phase transition}{
  name=Phase transition,
  text=phase transition,
  description={Here phase transition refers exclusively to the transition from a globally synchronous ecosystem to an asynchronous one, or vice-versa}
}

\newglossaryentry{Phase shifts}{
  name=Phase shifts,
  text=Phase shifts,
  description={Refers to the change in phase of oscillation of a single oscillator: from high population at even times to high population at odd times or vice-versa}
}

\newglossaryentry{spatiotemporal correlation}{
  name=Spatiotemporal correlation,
  text=spatiotemporal correlation,
  description={The Pearson correlation coefficient between the time series of the population first difference averaged over all pairs of sites that are the same distance apart}
}

\newglossaryentry{spatial correlation}{
  name=Spatial correlation,
  text=spatial correlation,
  description={The correlation coefficient of the two-cycle variable of sites separated by some distance at a single time}
}

\begin{frontmatter}
  \title{Long-range dispersal promotes spatial synchrony but reduces the length and time scales of synchronous fluctuations.}
  \author[1]{Davi Arrais Nobre\corref{cor1}}
  \ead{dv.nobr@gmail.com}
  \author[2]{Karen C. Abbott}
  \author[3,4]{Jonathan Machta}
  \author[5,4]{Alan Hastings}
  \cortext[cor1]{Corresponding author.}
  \address[1]{Department of Physics and Astronomy, University of California, Davis, One Shields Avenue, Davis, CA, 95616 USA.}
  \address[2]{Department of Biology, Case Western Reserve University, 10900 Euclid Ave, Cleveland, OH, 44106 USA.}
  \address[3]{Department of Physics, University of Massachusetts, Amherst, MA, 01003 USA.}
  \address[4]{Santa Fe Institute, 1399 Hyde Park Road, Santa Fe, NM, 87501 USA.}
  \address[5]{Department of Environmental Science and Policy, University of California, Davis, One Shields Avenue, Davis, CA, 95616 USA.}

  \begin{abstract}
Synchronous oscillations of spatially disjunct populations are widely observed in ecology. Even in the absence of spatially synchronized exogenous forces, metapopulations may synchronize via dispersal. For many species, most dispersal is local, but rare long-distance dispersal events also occur. While even small amounts of long-range dispersal are known to be important for processes like invasion and spatial spread rates, their potential influence on population synchrony is often overlooked, since local dispersal on its own can be strongly synchronizing. In this work, we investigate the effect of random, rare, long-range dispersal on the spatial synchrony of a metapopulation and find profound effects not only on synchrony but also on properties of the resulting spatial patterns. While controlling for the overall amount of emigration from each local subpopulation, we vary the fraction of dispersal that occurs locally (to nearest neighbors) versus globally (to random locations, irrespective of distance). Using a metric that measures the instantaneous level of global synchrony, we show that this form of long-range dispersal significantly favors the spatially synchronous state and homogenizes the population by decreasing the size of clusters of subpopulations that are out of phase with the rest of the metapopulation. Moreover, the addition of non-local dispersal significantly decreases the equilibration time of the metapopulation.
  \end{abstract}

 \end{frontmatter}

\section{\label{sec:intro}Introduction}
Spatially disjunct populations are known to display synchronous dynamics in different scenarios and over extensive distances \citep{liebhold2004, gouhier2010}. Such spatial synchrony can affect a population's stability, for example, by making it more susceptible to extinction, since the population density may be simultaneously low everywhere, or by dampening its temporal fluctuations, as the impact of local density dependence is moderated by regional processes \citep{abbott2011}. One mechanism behind long-range correlation is the Moran effect \citep{moran1953}, which drives populations over a wide range to a similar state by some shared exogenous factor, like the weather or instantaneous perturbations due to correlated noise \citep{stone1992, ranta1997-1, ostfeld2000, noble2018}. Another cause of spatial correlation is pairwise coupling between habitat patches through dispersal \citep{ranta1997-2, noble2015} or other interactions \citep{noble2018, shadi2020}.

Dispersal plays an important role in a species' dynamics and persistence \citep{adler1993, conradt2000, dispersal1-enc}. While it allows for colonization of new habitat patches and recovery from perturbations, it often leads to spatially correlated populations, increasing the probability of regional extinction \citep{reeve1990, adler1993, keeling2000}. Ultimately, the effect of dispersal in spatial synchrony depends on how dispersal rates are affected by factors such as local population density and distance \citep{abbott2011}. The vast majority of the work on metapopulations assumes one of three dispersal representations \citep{dispersal2-enc}: global dispersal, where each habitat patch is connected to all others and dispersal happens irrespective of distance \citep{roos1998, marti2003}, stepping-stone dispersal, where individuals migrate to nearest neighbors only \citep{durrett1994-1, araujo2008, noble2015}, and distance-dependent dispersal, where dispersing individuals are distributed to new patches depending on their starting position and a pre-defined dispersal kernel \citep{hastings1994, durrett1999, johst2002}.

Even stepping-stone interactions are enough to generate spatial synchrony over distances much greater than the dispersal distance and the correlation length of environmental noise in 2-dimensional habitats \citep{noble2015}. Empirical data and computer simulations help us understand the extent of such effect. Noble and colleagues showed that, even in the absence of the Moran effect, a pistachio orchard arranged on a rectangular lattice displayed long-range correlation consistent with a stepping-stone coupling \citep{noble2018}. In contrast, nearest neighbor dispersal alone is believed to be insufficient to create long-range synchrony in 1-dimensional habitats \citep{pathriabook}. However, the addition of occasional dispersal between sites chosen at random lead to the maintenance of spatial synchrony, both in a microcosm experiment \citep{fox2018} and in computer simulations \citep{kim2001}. Naturally, models that incorporate dispersal over greater distances tend to observe higher degrees of global synchrony \citep{hastings1994, roos1998, marti2003}.

The statistical method used to measure spatial synchrony may affect the conclusions drawn from data or models. Most studies use pairwise correlations of population abundance or rate of change to measure spatial synchrony  \citep{steen1996, liebhold2004, fox2018}. Phase synchrony has also been considered, but is typically also analyzed in a pairwise fashion \citep{cazelles2003}. While helpful, such measures may not always describe the global behavior of the metapopulation. Examples from statistical physics teach us that the range of pairwise correlations becomes arbitrarily large distances at a critical \gls{phase transition} \citep{pathriabook, Newmanbook}, in a phenomenon closely related to critical slowing down at a bifurcation. At such a critical point, long-range correlations exist, but the system is not globally synchronized. A global measure of synchrony is still lacking in most ecology literature thus far, but it could be useful to distinguish long-range correlated metapopulations from a globally synchronized ecosystem.

In this work, we study the effect of long-range dispersal on the spatial synchrony of a 2-D single species metapopulation model. We use a difference equation model to describe yearly reproduction patterns and study cyclic populations with period 2, which are widely observed in nature \citep{bjornstad1999, noble2018} and whose mathematical descriptions are well studied \citep{schaffer1986, hastings1993-2}. Dispersal is modeled by starting with a stepping stone dispersal and adding rare long-range dispersal, as we are interested in the effect of occasional distant migration, in an environment where most individuals migrate to nearest geographical neighbors, but some occasionally travel to distant habitat patches at random. To measure long-range synchrony, we introduce a global measure of spatial synchrony that takes into account both the phase and amplitude of oscillation of each population, and compare this metric with the mean pairwise correlation of the population dynamics, which is distance dependent. Previous work on this topic showed that when dispersal happens only to nearest neighbors the transition from a spatially asynchronous state to a spatially synchronous one is closely related to the well known Ising model in statistical physics \citep{noble2015}, which describes the transition between paramagnetic and ferromagnetic phases in some metals \citep{sole-phasetransitions, Newmanbook, pathriabook}. More recent work concluded that the introduction of random long-range dispersal, even if with low probability, makes the metapopulation qualitatively behave like a well mixed population with global dispersal \citep{davi2025}. Here we investigate further effects of the introduction of rare long-range coupling, focusing on the consequences for spatial synchrony.

\section{\label{sec:methods}Model and methods}

We use a discrete-time model of an oscillating population, with multiplicative log-normal noise to account for environmental fluctuations that are uncorrelated in space and time. The population density at a specific habitat patch $j$ at time $t+1$ can be written as 

\begin{equation}
    \label{eq:population}
    X_{j,t+1} = (1-z_j\epsilon)f(X_{j,t})e^{\xi_{j,t}}+\epsilon\sum_{k\in z_j} f(X_{k,t})e^{\xi_{k,t}},
\end{equation}
where $f(X_{j,t})$ is any deterministic density-dependent over-compensatory map in a period-2 cyclic regime \citep{noble2015, shadi2022}, $\epsilon$ is the fraction of individuals who disperse in each time step (and is thus a measure of the strength of the coupling between pairs of connected habitat patches), the summation is over the $z_j$ neighbors of patch $j$, and $\xi_{j,t}$ is a random normal variable with mean zero and variance $\sigma^2$. We call $\sigma$ the noise level throughout our work, and choose

\begin{equation}
    \label{eq:ricker}
    f(X_{j,t}) = X_{j,t}e^{r(1-X_{j,t})}
\end{equation}
which is the famous Ricker map \citep{ricker1954}. However, it is important to note that the functional form of $f(X_{j,t})$ does not qualitatively affect the results presented here as long as it is over-compensatory \citep{noble2015, pathriabook}.  That is, we use the Ricker model for illustration, but our results apply equally to other over-compensatory forms of local density dependence.

Local populations can interact via different mechanisms, which include but are not limited to direct migration, a dispersing predator \citep{ims2000}, pollination \citep{satake2000} and nutrient sharing via belowground connections such as root grafting \citep{prasad2017, shadi2020}. All forms of coupling are taken into account in the form of the coupling constant $\epsilon$, which we refer to as the dispersal rate. While it is reasonable to assume that dispersal will mostly happen to nearest neighboring sites, we should also expect a fraction of the dispersing population to reach further patches. We model that behavior by dynamically rewiring a regular square lattice, where each node is connected to its four nearest neighbors. At the beginning of each time step, before iterating Equation \eqref{eq:population}, we break each connection between nearest neighbors in the square lattice with probability $p$, replacing it with a new connection between one of those neighbors, chosen at random, and a random node on the network, provided that we do not recreate the broken connection nor create a connection that already exists. These long-range connections exist only during that time step, and the rewiring process is repeated at the beginning of every time step, always starting from the regular square lattice. This dynamic and temporary long-range dispersal can represent changing environmental conditions, like wind or water currents that can affect pollination or larvae dispersal, while the nearest neighbor connections account for active migration or resource sharing. Note that the average (Euclidean) dispersal distance is $1$ for $p=0$ and of order $p L$ for $p>0$ and large $L$, where $L$ is the linear size of the system. For large systems, the average dispersal distance increases significantly with the introduction of long-range dispersal, even if $p$ is fairly small.

The potential for correlated environmental perturbations to synchronize population dynamics has been studied in several different ways, ranging from the standard Moran effect -- in which the ongoing environmental noise ($\xi_{j,t}$ in our model) is spatially correlated \citep{moran1953} -- to the notion that more extreme events episodically bring populations into synchrony \citep[e.g.][]{ranta1999spatially,abbott2007does}.  Our approach in this study is most similar to the latter scenario, in that we study the model's behavior from synchronous initial conditions such as those that would be present shortly after a synchronizing event. However, because our focus is on the synchronizing potential of dispersal, we do not model additional environmentally-derived synchronizing mechanisms thereafter.  In other words, we account for synchrony due to a shared environment in our initial conditions, and then we study the role of dispersal in maintaining (or not) that synchrony through time.

\subsection{Spatial correlation of temporal dynamics}

A common way to determine spatial correlation in ecology is by calculating the correlation coefficient between the time series of two populations separated by some distance, and the decay of such correlation with distance gives the spatial scale of the correlations \citep{liebhold2004, buonaccorsi2001}. Different methods can be used to calculate these correlations, like using the Pearson or Spearman coefficients, applied to population densities or some transformed version of the abundance, or to local population growth rates \citep{bjornstad1999,buonaccorsi2001}. Taking the latter approach, here we look at the time series of the first difference in population densities,

\begin{equation}
    \label{eq:first-diff}
    \Delta_{j,t} = X_{j,t+1}-X_{j,t},
\end{equation}
and compute Pearson's correlation coefficient between $\Delta_{i,t}$ and $\Delta_{j,t}$ for all $i,j$ pairs.

Note that this is a pairwise measure, and therefore to get a measure of global synchrony, we average the correlations over all pairs of patches separated by the same distance. Averaged pairwise measures of correlation have been long used in ecology but are at best an approximation of the entirety of the population dynamics \citep{payne-enc}. Next, we introduce a global measure of spatial synchrony.

\subsection{Synchronization order parameter}

\begin{figure}[!ht]
\centering
\includegraphics[scale=0.23]{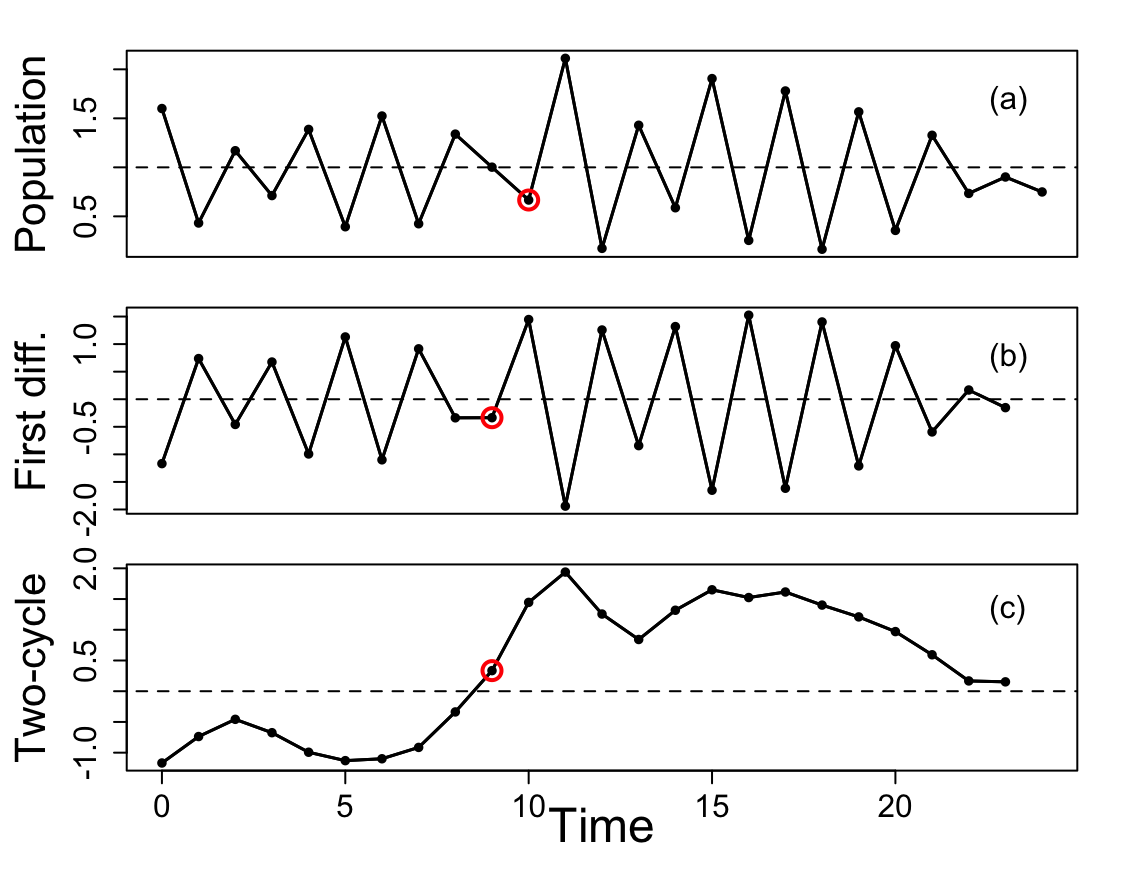}
\caption{\label{fig:time} Shown in (a) is the time series of the population abundance according to the noisy Ricker map defined in Equations \eqref{eq:population} and \eqref{eq:ricker}. The population oscillates with period 2 around the environmental carrying capacity ($X_{j,t}=1$, the dashed line) and noise can cause it to change the phase of oscillation (red circle). In (b), the population first difference, Equation \eqref{eq:first-diff}, of the population abundance shown in (a). The dashed line shows indicates where zero is, and for an oscillating population the first difference is expected to change sign at every time step. \gls{Phase shifts} can be identified by two consecutive points with the same sign. Finally, (c) shows the two-cycle variable for the same data, which is expected to have the same sign for an oscillating population, and phase shifts can be identified by a change in sign.}
\end{figure}

\cite{noble2015} took inspiration from statistical physics to define the so called synchronization order parameter, which provides a global measure of synchrony based on the entire metapopulation, rather than the set of all pairs. First, we define a two-cycle variable that is similar to the first difference in population abundance, Equation \eqref{eq:first-diff}, but includes an alternating sign term so that it measures the phase of oscillation, and not the instantaneous change in population. The two-cycle variable of habitat patch $j$ at time $t$ is given by 

\begin{equation}
    \label{eq:two-cycle}
    m_{j,t} = (-1)^{t}(X_{j,t+1}-X_{j,t})=(-1)^{t}\Delta_{j,t}.
\end{equation}
Figure \ref{fig:time} shows the time series of the population abundance, $X_{j,t}$, the respective population first difference, $\Delta_{j,t}$, and two-cycle variable, $m_{j,t}$. 

Averaging the two-cycle variable over all habitat patches gives us the instantaneous synchronization order parameter, Equation \eqref{eq:order}, which is a measure of global synchrony that can be obtained with data from only two time steps,

\begin{equation}
    \label{eq:order}
    m_t = \left|\frac{1}{N}\sum_j m_{j,t}\right|.
\end{equation}

This definition has been successfully used to describe a phase transition between synchrony and asynchrony in systems of ecological oscillators \citep{noble2015, noble2018, shadi2022, davi2025}. In the spatially asynchronous state, where roughly half of the habitat patches are in each of the two possible phases of oscillation, it is close to zero and the two-cycle variable has very short spatial correlations. In the synchronous state, most of the habitat patches are in the same phase and the synchronization order parameter is nonzero, but deviations in population first difference also have a very short spatial correlation. At the critical point (a threshold combination of coupling strength and noise) where the phase transition happens, the instantaneous order parameter is still expected to be close to zero for large systems, but it displays large fluctuations and therefore we typically see a scale-free pattern, with spatial correlations that span the length scale of the system \citep{pathriabook, noble2015}. 

It is important to distinguish the spatial correlation of such deviations from the pairwise spatiotemporal correlation of the population time series that is usually measured in ecology. The \gls{spatiotemporal correlation} is more similar to the instantaneous order parameter in the sense that it remains nonzero at large distances. The instantaneous \gls{spatial correlation} measures the length scale of deviations in the two-cycle variable, which is expected to be small in both the synchronous and asynchronous states, reaching a maximum at the transition point. To measure the instantaneous spatial correlation, we use the connected correlation function from statistical physics \citep{pathriabook, grigera2021}, which is an unnormalized version of the Pearson correlation coefficient of the instantaneous two-cycle variable in space, and at time $t$ is given by 

\begin{equation}
    \label{eq:correlation-function}
    G_t(|\mathbf{r}|) = \sum_j (m_{j,t}-m_t)(m_{j+\mathbf{r},t} - m_t),
\end{equation}
where the the distance between habitat patches on a regular lattice is simply $|\mathbf{r}|=\sqrt{\Delta x^2+\Delta y^2}$, where $\Delta x$ and $\Delta y$ are the Euclidean horizontal and vertical distances, respectively, and go up to $L/2$ due to the periodic boundary conditions. The connected correlation function decays exponentially at long distances, and such exponential decay is described by the fluctuation correlation length, which corresponds to the expected length scale of the clusters of subpopulations whose two-cycle variables deviate from the average value in a correlated fashion. The fluctuation correlation length can be obtained by fitting the Fourier transform of $G_t(|\mathbf{r}|)$ to a Lorentzian curve for low wavenumbers \citep{jankebook, janke1994, grigera2021}.

From now on, we use the word spatiotemporal correlation to refer to the pairwise measure of correlation of the time series of the first difference in population abundance, and spatial correlation to refer to Equation \eqref{eq:correlation-function}.

\subsection{Simulation details}

In all of our work, we use the Ricker growth rate (see Equation \eqref{eq:ricker}) $r=2.3$, which is close to the super-stable period-2 cycle of the Ricker map, and a dispersal rate $\epsilon = 0.025$. We performed simulations on an $L\times L$ square lattice for system sizes $L=16, 32, 64,$ and $128$, using 12 environmental noise levels, $\sigma$, evenly spaced in the interval $[0.14, 0.195]$, and 13 values of long-range dispersal fraction including eleven evenly spaced values in the interval $[0, 0.2]$ and the higher values $p=0.6$ and $1$. The initial condition was chosen to be in a highly spatially synchronous state, with the initial population ($t=0$) at each site chosen at random from a normal distribution with mean $1.6$, which is close to the high population value in the 2-cycle with $r=2.3$, and standard deviation $0.2$. Such initial condition simulates the outcome of a strong globally synchronizing event, such as a long-range climatic forcing \citep{prost2002}. The simulation follows with no correlated extrinsic forcing, and dispersal being the only mechanism available to maintain synchrony. We perform $10^6$ time steps according to Equation \eqref{eq:population}, which ensures systems for all choices of parameters have enough time to reach the asymptotic behavior. In our results, we present measures for both the beginning of the time series, considering the first $1000$ times steps following the globally synchronizing event, and the asymptotic behavior, which can be reached in time scales that range from a few dozen time steps to several thousands. For asymptotic behavior measurements, we considered the final $1000$ time steps of the simulation. The next section presents such results, which focus on the effect of the fraction of long-range dispersal, $p$, in maintaining global spatial synchrony.

\section{\label{sec:results}Results and discussion}

\begin{figure}[!ht]
\centering
\includegraphics[width=0.9\linewidth]{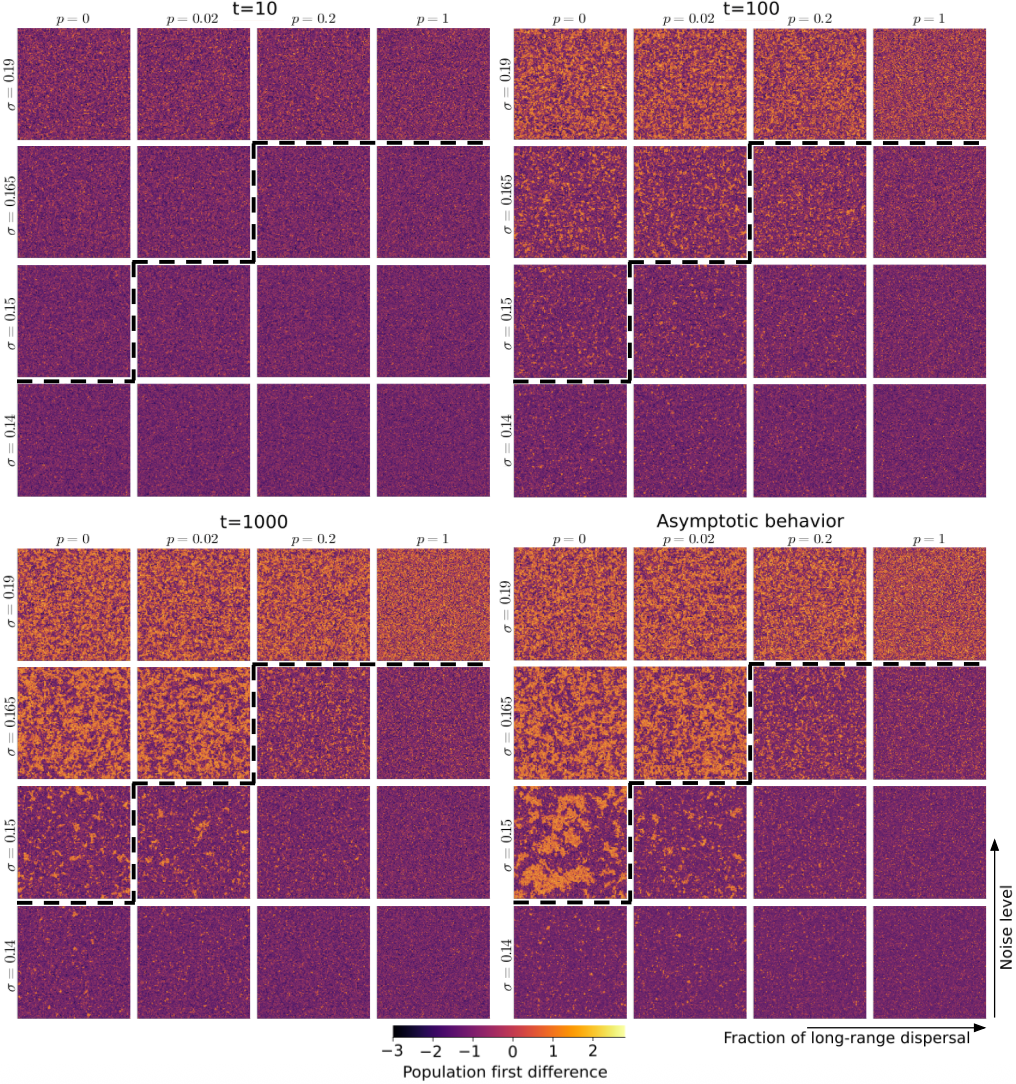}
\caption{\label{fig:snapshots} Snapshots of the population first difference (Equation \eqref{eq:first-diff}, or equivalently the two-cycle variable (Equation \eqref{eq:two-cycle}) show the effect of long-range dispersal in maintaining synchrony. For a relatively low noise level, $\sigma =0.14$, all populations remain synchronous, while for high noise, $\sigma = 0.19$, all populations become asynchronous. For intermediate noise values, some populations remain synchronous depending on the value of $p$. The initial condition was high population at $t=0$, which is the phase of oscillation represented by the purple color in the snapshots. Note that snapshots with the same $p$ and $\sigma$ values are from the same run.}
\end{figure}

\begin{figure}[ht]
\centering
\includegraphics[scale=0.23]{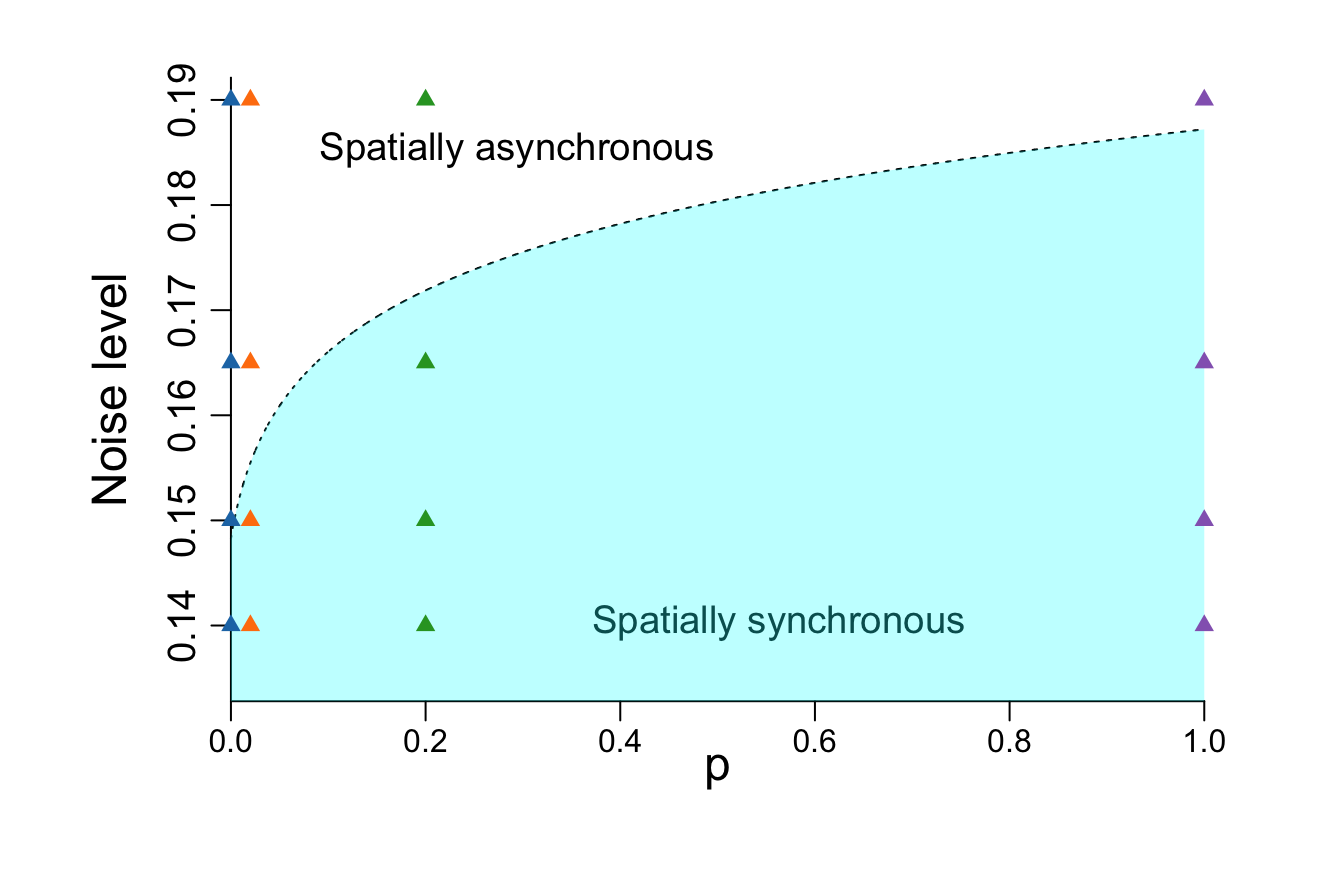}
\caption{\label{fig:phase-diagram} Phase diagram showing the critical line that separates the spatially synchronous and the spatially asynchronous states. The triangles represent the 16 snapshot rows in Figure \ref{fig:snapshots}.}
\end{figure}

Our results agree with the previously observed conclusion that increasing the fraction of long-range dispersal favors the spatially synchronous state \citep{davi2025}. However, the spatial patterns observed can be strongly affected by the presence of long-range dispersal. Figure \ref{fig:snapshots} shows the snapshots of the metapopulation at different times for systems of size $L=128$. The color indicates the first difference in population abundance; notice that since each snapshot is shown at a fixed time, Equations \eqref{eq:first-diff} and \eqref{eq:two-cycle} are equivalent. Each quadrant shows a different simulation time: $t=10$, very close to the globally synchronous initial state, $t=100$, $t=1000$, and asymptotic behavior, shown using the final state after one million time steps. Within each quadrant, each row has a constant noise value, which decreases from top to bottom and the values chosen were $\sigma=0.19, 0.165, 0.15,$ and $0.14$, and each column represents a different fraction of long-range dispersal, the first being $p=0$, the stepping stone case, the last $p=1$, which is completely random distance-independent dispersal, and the two intermediate cases $p=0.02$ and $p=0.2$ to depict the effect of a small fraction of long-range coupling. The dashed black line between snapshots shows the separation between parameter values that correspond to a spatially synchronous (below the line, i.e. with smaller noise and more long-range dispersal) or asynchronous (above the line) asymptotic behavior, which is shown in more detail in Figure \ref{fig:phase-diagram}.

From the snapshots, we can observe both global synchrony (where more overall similarity in color indicates higher synchrony) and spatial pattern (i.e., whether like colors are clumped or scattered across the lattice). We see that all populations remain synchronous at a low noise level, with a few patches of populations (orange in these snapshots) in the opposite phase of oscillation arising due to noise, especially for the $p=0$ case, and all populations seem to reach the asymptotic behavior state within a few hundred time steps. For $\sigma=0.15$, we see larger clusters of the different phase arising in the cases $p=0$ and $0.02$, but the asymptotic behavior snapshot for $p=0$ shows a fractal pattern with synchronous clusters of subpopulations in both phases of oscillations at all length scales, while non-zero values of $p$ remain highly synchronous. For $\sigma=0.165$, the two lowest values of $p$ show very similar spatially asynchronous patterns, with synchronous clusters spanning small but significant length scales, and the two highest values of $p$ remain synchronous, but with a higher presence of asynchronous sites that are very scattered and do not clump over significant length scales. Finally, at high noise, all populations become asynchronous, but again there is some spatial structure at low values of $p$, while higher values of $p$ shows sites at different phases of oscillation that are scattered and do not form significant synchronous clusters. 

The maintenance or destruction of global synchrony can be summarized by a phase diagram. For each value of $p$, there is a critical noise level that separates the synchronous and asynchronous states. Determining the critical noise requires methods from statistical physics that are outside the scope of this report, but is presented in our earlier work \citep{davi2025}. Figure \ref{fig:phase-diagram} shows the phase diagram, where the dashed line was fitted through the critical points calculated for 13 different values of $p$ \citep{davi2025}. The triangles depict the 16 combinations of $p$ and $\sigma$ shown on the snapshots in Figure \ref{fig:snapshots} and throughout this paper. Note that for $\sigma =0.14$ all triangles are in the spatially synchronous region, which means that synchrony is maintained, although the degree of synchrony depends on how close the point is from the critical line that separates the two regions. For $\sigma=0.15$, the $p=0$ point is in the spatially asynchronous region, but close to the critical line, which explains the fractal pattern observed in the asymptotic behavior. Such proximity to the critical line also explains why the system takes a long time to reach the asymptotic behavior, an effect that is seen on the snapshots and is a consequence of critical slowing down \citep{sole-phasetransitions, davi2025}. The phase diagram shows how introducing long-range dispersal favors spatial synchrony, and the change in critical noise level is greater at low values of $p$. 

\begin{figure}[ht]
\centering
\includegraphics[width=1\linewidth]{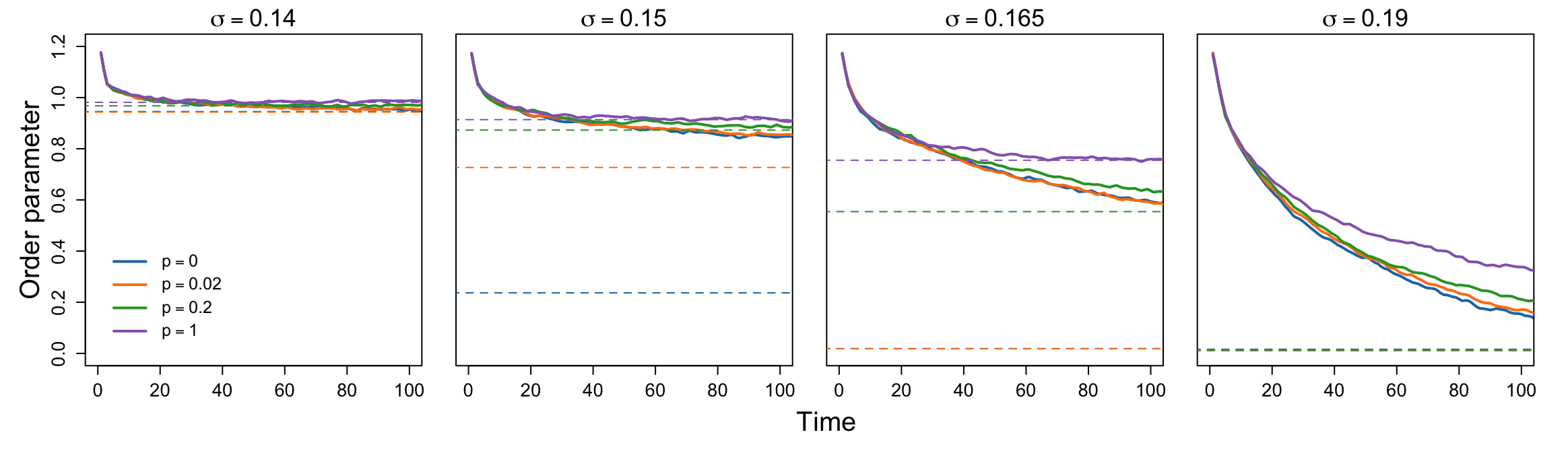}
\caption{\label{fig:order-time} Order parameter time series for the first 100 time steps of the simulation. Note that all time series start in a highly synchronous value, which is a representation of the globally synchronizing event at $t=0$. The dashed lines correspond to the asymptotic value of the order parameter, the average of $m_t$ over the final 1000 time steps of the simulation. For $\sigma=0.14$, all metapopulations are in the synchronous state, while for $\sigma=0.15$ the $p=0$ system is in the asynchronous state, for $p=0.165$ both $p=0$ and $0.02$ are in the asynchronous state, and for $\sigma=0.19$ all systems go out of synchrony, which is in accordance with the points shown in Figure \ref{fig:phase-diagram}.}
\end{figure}

Another way to observe how spatial synchrony changes through time is to look at the instantaneous synchronization order parameter, Equation \eqref{eq:order}, or the spatiotemporal correlation of the population first difference as a function of distance. Naturally, these quantities are time dependent, and will typically decrease after a globally synchronizing event. Figure \ref{fig:order-time} shows the time series of the order parameter over the first 100 time steps of the simulation (thick solid lines), and the dashed lines indicate the asymptotic value, determined by the time average of the order parameter in the final 1000 time steps. We see results consistent with the snapshots. For $\sigma=0.14$ all metapopulations reach the synchronous asymptotic behavior within the first 100 time steps. For $\sigma=0.15$, $p=0$ approaches a disordered state very slowly, taking much more than 100 time steps to reach it due to critical slowing down. If one were to observe the system in those early stages they could be led to the incorrect conclusion that dispersal is maintaining spatial synchrony, but that's simply a long time consequence of the global synchrony introduced by the Moran effect at $t=0$. Note that at $\sigma=0.15$, the metapopulation with $p=0.02$ also does not reach the asymptotic behavior within the first 100 time steps, but, as shown in the snapshots, asymptotic behavior is reached after about 1000 updates, and the longer equilibration time is again a consequence of the proximity to the critical line. For $\sigma=0.165$, the metapopulations with $p=0$ and $0.02$ are both deeply into the asynchronous phase, and don't have time to get to the asymptotic behavior within 100 time steps, but their time series show a clear negative slope at $t=100$, such that even short time observation of the system is sufficient to understand that dispersal is not maintaining synchrony. Finally, for $\sigma = 0.19$, all metapopulations go out of synchrony and this result is easily observed for all systems, even if the value of the order parameter does not reach zero within 100 time steps. All these results are visually confirmed by the snapshots in Figure \ref{fig:snapshots}. It is interesting to notice that, even though $\sigma=0.19$ is close to the critical line for $p=1$ (see figure \ref{fig:phase-diagram}), the metapopulation still reaches the asymptotic behavior in a fairly short time compared to the metapopulations with $p=0$ and $\sigma=0.15$, and $p=0.02$ and $\sigma=0.165$. This shorter equilibration time is an effect of the better mixing of the population with $p=1$, and has been rigorously measured in our previous work \citep{davi2025}.

\begin{figure}[ht]
\centering
\includegraphics[width=1\linewidth]{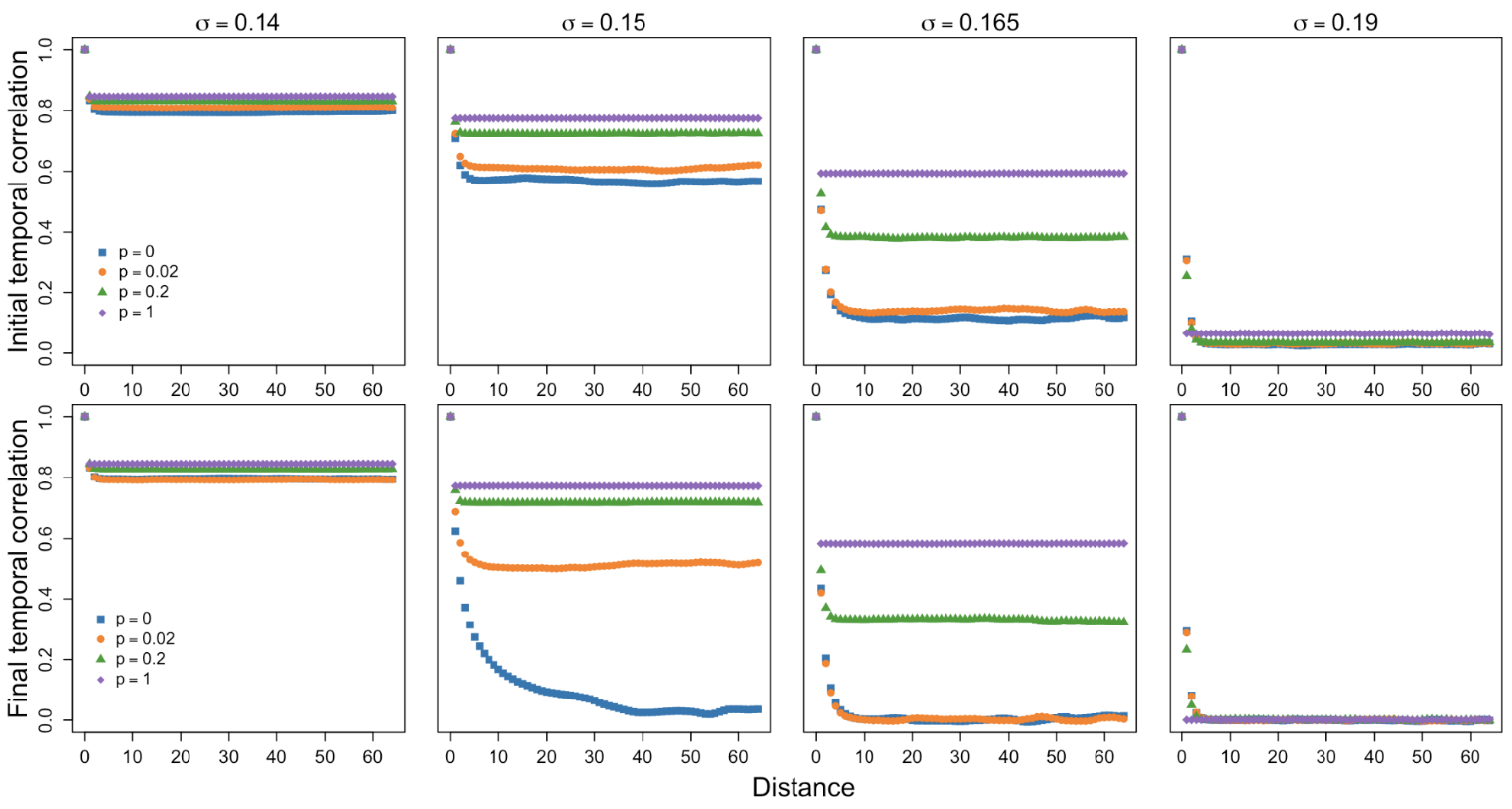}
\caption{\label{fig:time-correlations} Correlation of the time series of first difference in population abundances at sites separated by a given distance. Horizontal and vertical Euclidean distances were considered. The top row shows the correlation in the first 1000 time steps of the simulation, and the bottom row shows the correlation in the last 1000 time steps, after the asymptotic behavior was reached.}
\end{figure}

The traditional way to determine long-range spatial synchrony in ecology is by measuring the correlation coefficient between the time series of two disjunct populations. Figure \ref{fig:time-correlations} shows this measure for the population first difference, Equation \eqref{eq:first-diff}, for two different time intervals: the top plots are the spatiotemporal correlation over the first 1000 time steps of the simulation, and the bottom plots are for the final 1000 time steps, when asymptotic behavior has been reached. We show the spatiotemporal correlation as a function of distance, and each point is the average of all pairs of sites at a given distance considering horizontal and vertical Euclidean distances for metapopulations with $L=128$. The conclusion drawn from these plots are similar to the ones from the time series of the synchronization order parameter, with the most striking effect being the difference in the spatiotemporal correlation for the $p=0$ metapopulation when $\sigma=0.15$. Note, however, that this method is highly sensitive to which time interval we choose. For the $\sigma=0.165$ plot, looking at the spatiotemporal correlation over the first 100 time steps instead of the first 1000 would show us that the $p=0$ and $p=0.02$ systems remain highly synchronous, as shown in Figures \ref{fig:snapshots} and \ref{fig:order-time}, but we would not have the temporal trend of the correlation to guide us towards a correct conclusion, like we do in Figure \ref{fig:order-time}, where the slope of the curve helps us determine if the asymptotic behavior has been reached.

Overall, both methods appear to have a similar efficacy, but provide different pieces of information. The spatiotemporal correlation graphs allow us to determine the length scale of the spatial synchrony, such that if the average correlation coefficient remains high for all distances, the entire system is mostly synchronous (see the snapshots for $\sigma=0.14$ in Figure \ref{fig:snapshots}). If the correlation coefficient decays to zero, the length of the decay informs us about the average size of the synchronous subpopulations clusters (see the snapshots with $\sigma=0.15$ and $p=0$ in asymptotic behavior, and the asymptotic snapshots in the asynchronous phase, with synchronous clusters with lengths similar to the distance at which the spatiotemporal correlation becomes zero at the bottom plots in Figure \ref{fig:time-correlations}). Note that, for $p=1$, the spatiotemporal correlation has a constant value for all non-zero distances, since that corresponds to completely random dispersal with no spatial structure. This can also be seen in the snapshots, where sites at the same phase are scattered across the network without forming any spatial structure. On the other hand, spatiotemporal correlations do not provide much information about the time evolution of the system, and may lead to incorrect conclusions about the ability of dispersal to maintain synchrony depending on which time interval we consider. For that purpose, the time series of the order parameter is a better measure, since it shows a clear trend of the global synchronization of the metapopulation, making it easier to distinguish between long unstable transients and the asymptotic behavior. When we have a set of parameters very close to criticality, in this work the $p=0$ system with $\sigma=0.15$, we have quasi-stable states that have a very long lifetime. These correspond to long transients that are known to occur in spatially structured population models \citep{hastings1994}. In such cases, both methods may lead to the wrong conclusion about the maintenance of spatial synchrony due to dispersal, since to reach the asymptotic behavior the system would take many more generations than are ecologically possible. However, the lifetime of these quasi-stable states significantly decreases as we increase $p$, even if we remain close to the critical line \citep{davi2025}.

\begin{figure}[ht]
\centering
\includegraphics[width=1\linewidth]{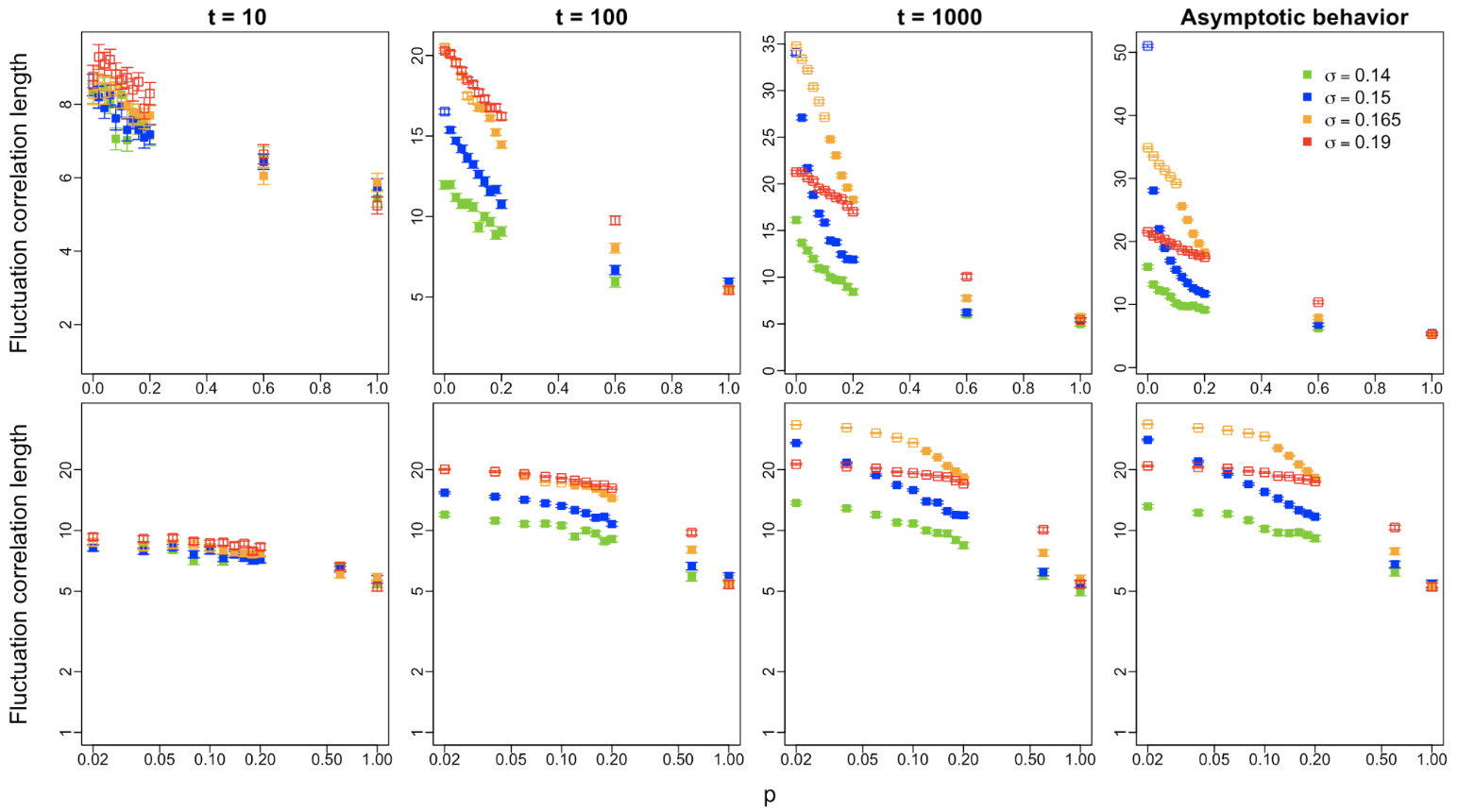}
\caption{\label{fig:correlation-length} Fluctuation correlation length as a function of $p$ for the different noise levels shown throughout this report at the times shown in the snapshots in Figure \ref{fig:snapshots}. The lower plot shows a subset of the points in the upper plot in a log-log scale. Solid symbols indicate that the combination of parameters for the point is in the spatially synchronous state, while open symbols indicate the point is in the asynchronous state. The error bars for times $t=10, 100,$ and $1000$ were calculated from the standard deviation of the snapshots from 100 independent runs with 1000 time steps each. The asymptotic behavior panels were plotted using data from a single run, but using 100 different snapshots in the asymptotic behavior spaced by 5000 time steps.}
\end{figure}

One more point that can be observed in the snapshots in Figure \ref{fig:snapshots} is that, even in the spatially synchronous states, we observe small clusters of synchronous subpopulations in the opposite phase of oscillation from the majority of the metapopulation. These sites are responsible for decreasing the order parameter as seen in Figure \ref{fig:order-time}, or, similarly, decreasing the correlation coefficient as seen in Figure \ref{fig:time-correlations}. The fluctuation correlation length measures the average size of these clusters. Note that it differs from the usual definition of correlation length in ecology in that, typically, the correlation length in ecology would simply be the system size in the synchronous state, while the fluctuation correlation length measures the size of the synchronous clusters that are negatively correlated to the rest of the sites. Figure \ref{fig:correlation-length} shows our measurements for the fluctuation correlation length as a function of $p$ over time, which provides insights on how subpopulations that are out of phase with most of the ecosystem clump together.

At $t=10$, we see that there is very little spatial structure, with a small dependence of the fluctuation correlation length with $p$, and only small clusters of populations that deviate from the average. At the beginning of the simulation, $t=0$, the population at each site was chosen irrespective of spatial location, which means that the fluctuation correlation length should be close to zero. The small values observed at the earliest time shown in Figure \ref{fig:correlation-length} are a consequence of that, since dispersal still had not had time to create large clusters that deviate from the dynamics introduced at $t=0$. As the simulation time increases, we see that spatial structure arises. At $t=100$, we already see a clear dependence of the cluster size with the fraction of long-range dispersal, and at $t=1000$ all points with $p>0$ are very close to their asymptotic value. However, $p=0$ takes a longer time to reach the asymptotic behavior, especially for $\sigma=0.15$, which is consistent will all our previous observations. In general, we see that increasing the fraction of long-range dispersal significantly decreases the size of synchronous deviating clusters. This result is intuitive if we think that the $p=1$ case corresponds to the absence of spatial structure, and therefore leads to the most scattered two-cycle variable distribution on the network. A less intuitive result seen in Figure \ref{fig:correlation-length} is that the $\sigma=0.14$ case has the smallest fluctuation correlation length, which shows that deviations in the spatially synchronous state tend to be smaller than in the asynchronous state. While this can be attributed to the proximity to the critical line for high values of $p$, it remains true even for $p=0$, for which $\sigma=0.14$ is much closer to the critical line than $\sigma=0.19$ and still has a smaller fluctuation correlation length. This is consistent with the snapshots in Figure \ref{fig:snapshots}, and teaches us that the way the ecosystem approaches the asymptotic behavior is different depending on where the parameters are in the phase diagram (Figure \ref{fig:phase-diagram}). 

\section{\label{sec:conclusion}Conclusion}

Our work confirms the widely known result that dispersal alone is able to maintain synchrony, as long as the environmental noise is below some threshold \citep{noble2015}. However, this threshold can be significantly larger if there is long-range dispersal \citep{davi2025}.  

We showed that a typical measure of spatial synchrony in ecology, the correlation coefficient between time series of population first differences, provides valuable, but incomplete, insights on the state of the metapopulation. Our additional analyses focused on slightly modifying the definition of the population first difference to include a factor of $(-1)^t$, defining a two-cycle variable that only changes sign when the corresponding subpopulation changes its phase of oscillation. Averaging this two-cycle variable over the entire metapopulation gives us the instantaneous order parameter, which is a global measure of population synchrony, not a pairwise approximation. These quantities have been used in recent literature \citep{noble2015, noble2018, shadi2022, vahini2020, davi2025}, and their behavior over space and time typically provides a better method to draw conclusions regarding spatial correlations and the state of the metapopulation. 

Our results show that, even if the fraction of dispersing individuals remains the same, occasional random long-range dispersal significantly favors spatial synchrony and population homogeneity, decreasing the typical size of subpopulation clusters that deviate from the rest of the metapopulation. Furthermore, a small fraction of long-range dispersal is enough to significantly reduce the time the system takes to reach the asymptotic behavior. However, dispersal alone will typically take many decades, and in some cases centuries, to bring the ecosystem to its asymptotic behavior. Thus, large-scale events with the potential to synchronize the metapopulation are remarkably relevant, even if very rare. Consequently, the state at which an ecosystem is observed  likely depends on  events that may have occurred long before the observation time.

\printglossary

\bibliography{apsbib}
\end{document}